\documentclass[12pt]{amsart}
\usepackage{amsbsy,amssymb,amscd,amsfonts,latexsym,amstext,delarray,
amsmath,graphicx} 
\input xypic

\usepackage{color}

\numberwithin{equation}{section}

\def\bO{{\mathbb O}}

\def\C{{\mathbb C}}
\def\F{{\mathbb F}}
\renewcommand{\H}{{\mathbb H}}
\def\N{{\mathbb N}}
\renewcommand{\P}{{\mathbb P}}

\def\R{{\mathbb R}}
\def\Z{{\mathbb Z}}

\def\cH{{\mathcal H}}

\def\cK{{\mathcal K}}

\def\cN{{\mathcal N}}

\def\Conf{{\rm Conf}}

\title{Anyons in Geometric Models of Matter}
\author{Michael Atiyah and Matilde Marcolli}
\address{School of Mathematics, The University of Edinburgh, Edinburgh
EH9 3FD Scotland, UK}
\email{M.Atiyah@ed.ac.uk}
\address{Division of Physics, Mathematics, and Astronomy, California Institute of Technology, 1200 E California Blvd, Pasadena, CA 91125, USA}
\email{matilde@caltech.edu}

\begin{document}
\maketitle

\begin{abstract}
We show that the ``geometric models of matter" approach proposed by the first
author can be used to construct models of anyon quasiparticles with fractional
quantum numbers, using $4$-dimensional edge-cone 
orbifold geometries with orbifold singularities along embedded $2$-dimensional
surfaces. The anyon states arise through the braid representation of surface braids
wrapped around the orbifold singularities, coming from multisections of the
orbifold normal bundle of the embedded surface. We show that the resulting braid
representations can give rise to a universal quantum computer.
\end{abstract}

\section{Introduction}

The main purpose of this paper is to explore new aspects of the
geometric approach to models of matter introduced and developed by the first author and collaborators 
in \cite{AFS}, \cite{AtMan}, \cite{AMS}, \cite{FraMa}. 
All the mathematical facts we refer to in this 
paper are known and the reader will be able to find more information about them in 
the extensive list of references that we provide. The only novelty we introduce is the
observation that, within these geometric
models of matter, it is possible to describe systems of anyon quasi-particles
with fractional quantum numbers, based on orbifold geometries. We show 
that these geometries allow for the presence of surface
braids, wrapped around the $2$-dimensional orbifold singularities, and that these surface
braids give rise to associated braid representations that determine anyon states.
We also show that the anyons that arise from these surface braid configurations can 
behave like a universal quantum computer.

\smallskip

\section{Orbifolds and geometric models of matter}\label{orbiSec}

It was shown in \cite{AMS} that certain classes of $4$-dimensional Riemannian manifolds
with self-dual Weyl tensor behave in many ways like elementary particles, and can be used
to provide geometric models of matter. These manifolds include gravitational instantons
like the Taub-NUT manifold \cite{NUT}, \cite{Taub} or the Atiyah--Hitchin manifold \cite{AH},
as well as compact manifolds like $\C\P^2$ or $S^4$. These were, respectively, proposed in \cite{AMS} as
models for the proton, the electron, the neutron, and the neutrino. More recent ongoing
developments of these geometric models of matter have assigned a different interpretation
to some of these manifolds, as we discuss briefly below. These static models of matter were 
made dynamical in \cite{AFS}, by considering $(4+1)$-dimensional Ricci-flat spacetimes
describing evolving Taub-NUT geometries. Models of systems of charged particles, based 
on gravitational instantons of types $A_k$ and $D_k$ were constructed in \cite{FraMa}.
This approach can be viewed, to some extent, as a geometrization 
of Skyrmion models.

\smallskip

The geometric models of matter originally proposed in \cite{AMS} have   
more recently been extended to models of nuclear physics and beta decay, in
work of the first author and Nick Manton \cite{AtMan}, using algebraic surfaces as geometric models of nuclei,
with lepton and baryon numbers related to the topological invariants $c_2$ and $c_1^2$
and to the Enriques-Kodaira classification of compact complex surfaces. In this setting,  
beta decay is related to blow-up operations, and the ``valley of stability" is realized by 
the zero signature region where $c_1^2 = 2c_2$. More general models involve 
non-self-dual cases. In these geometric realizations of atomic
level physics, the constituent elementary particles can be obtained when pulling the compact 
manifold apart by stretching a long neck. 

\smallskip

In the present paper, we focus on a different aspect of these geometric models of matter, 
namely we construct geometric models of systems of quasi-particles, 
based on $4$-dimensional orbifold geometries studied by the first author 
and LeBrun in \cite{AtLe}.

\smallskip

In other recent work of the first author, \cite{AtS6}, related to the famous question on the
existence of complex structures on the sphere $S^6$, odd and even modules for the
quaternion group of order eight are considered, where the odd modules are 
faithful quaternionic representations, with value $-1$ on the center, while
the even ones descend to abelian modules and have value $+1$ on the
center. We expect that these odd and even types will also play a role in the geometric
models of matter, where they may be related to topological insulators.
We plan to investigate further possible connections to topological insulators and  
quantum computing aspects of the present work.

\smallskip

\subsection{Edge-cone metrics and orbifolds}\label{edgeconeSec}

Let $M$ be a smooth compact $4$-dimensional manifold and $\Sigma$ a smoothly
embedded compact $2$-dimensional surface. According to \cite{AtLe}, \cite{LeBrun},
an edge-cone metric on $(M,\Sigma)$ with cone angle $2\pi \beta$, for some $\beta\in \R^*_+$, 
is a smooth Riemannian metric on $M\smallsetminus \Sigma$ that is modelled on a 
$2$-dimensional cone in the
directions transversal to $\Sigma$, while it is smooth in the directions parallel to $\Sigma$.
More precisely, in a system $(\rho,\theta,x^1,x^2)$ of transversal polar coordinates near $\Sigma$, 
such a metric can be written in the form 
\begin{equation}\label{edgecone}
g = d\rho^2 +\beta^2 \rho^2 ( d\theta + u_j dx^j)^2 + w_{ij} dx^i dx^j + \rho^{1+\epsilon} h, 
\end{equation}
where $h$ is a symmetric tensor that has continuous derivatives of all orders with respect to
vector fields with vanishing normal component along $\Sigma$ (infinite conormal regularity).

\smallskip

The type of $4$-dimensional geometries we are especially interested in here are those
that arise as orbifolds. This means a compact $4$-dimensional $M$ that admits an atlas 
of local uniformizing charts $U_\alpha$ that are homeomorphic $U_\alpha \simeq V_\alpha/G_\alpha$ 
to quotients of open sets $V_\alpha\subset \R^4$ by finite groups $G_\alpha$. An orbifold
$M$ decomposes $M=M_{sing}\cup M_{reg}$ as disjoint union of the set of singular (orbifold) 
points and the complementary set of regular points.
Orbifold points are fixed points of the stabilizer group $G_\alpha$ in some chart $U_\alpha$.
In particular, we consider orbifolds where the set of singular points
$M_{sing} =\Sigma$ is a $2$-dimensional embedded surface. 
An orbifold $M$ is a good orbifold if it is a global orbifold quotient, namely there is a smooth 
$4$-manifold $X$ and a finite group $G$ such that $M=X/G$. 

\smallskip

Consider an orbifold $(M,\Sigma)$ where $M_{sing}=\Sigma$ is an embedded
surface, and where the orbifold structure near $\Sigma$ can be described in a local
uniformizing chart as a quotient $\C^2/G_\nu$, where $G_\nu$ is the cyclic group 
$G_\nu=\Z/\nu\Z$, for some $\nu\geq 2$, with the generator acting by 
$(w,\zeta)\mapsto (w, e^{2\pi i/\nu} \zeta)$.
An orbifold edge-cone metric on such an orbifold $(M,\Sigma)$, 
with cone angle $2\pi \beta$ and $\beta=1/\nu$, is an edge-cone metric on the smooth 
$4$-manifold $M\smallsetminus \Sigma$, that is represented in a local chart as a 
$\Z/\nu\Z$-invariant metric.

\smallskip

A result of \cite{LeBrun} shows that if an edge-cone metric with $\beta =1/\nu$ 
is an Einstein metric, then it must be an orbifold Einstein metric. Topological obstructions
to the existence of Einstein edge-cone metrics were obtained in \cite{AtLe}, in terms of index
calculations of signature and Euler characteristic. Smooth obstructions to the existence
of Einstein edge-cone metrics were then obtained in \cite{LeBrun} using Seiberg--Witten theory.

\subsection{Gravitational instantons}\label{gravinstSec}

Several examples of $4$-dimensional geometries with Einstein edge-cone metrics
considered in \cite{AtLe}, \cite{LeBrun}, \cite{LoVia} also happen to be self-dual or
anti-self-dual. We refer to $4$-dimensional Riemannian geometries that are both
self-dual (or anti-self-dual) and Einstein as {\em gravitational instantons}. 

\smallskip

A self-dual $4$-manifold $M$ has a twistor space $Z=Z(M)$, which is a $3$-dimensional 
complex manifold that fibers over $M$ with $\C\P^1$ fibers. This provides an important connection
between Riemannian and complex geometry. Moreover, self-dual $4$-manifolds behave well
under connected sums. Indeed, as shown in \cite{DoFri}, the existence of a self-dual metric
on a connected sum of two self-dual $4$-manifolds, which is close to the original ones outside 
a neck where the connected sum is performed, can be formulated in terms of  twistor geometry.
It results from the existence of a deformation of a singular complex $3$-manifold obtained by 
blowing up the twistor spaces $Z_i=Z(M_i)$ along a $\C\P^1$ fiber of the fibration to $M_i$ and
identifying the exceptional divisors $E_i$ of the blowups $\tilde Z_i$. 
It is shown in \cite{DoFri} that if the singular space $\tilde Z=\tilde Z_1 \cup_{E_1\simeq E_2} \tilde Z_2$
admits a smooth Kodaira--Spencer--Kuranishi deformation $Z$, then $Z$ is in fact the twistor space $Z(M)$
of a self-dual structure on the connected sum $M=M_1\# M_2$. 

\smallskip

This connected sum property of self-dual metrics will be useful in the following, to construct 
systems of quasi-particles. However, for other aspects of the construction we discuss in this 
paper, it will not be necessary to assume that the $4$-dimensional geometries we consider 
are necessarily gravitational instantons, and it is possible to relax this condition 
for our purposes.

\subsection{Dynamical models}\label{dynmodSec}

As mentioned above, the geometric models of matter described in \cite{AMS} in terms of gravitational
instantons, given by $4$-dimensional self-dual Riemannian manifolds with an Einstein
metric, possibly with edge-cone structure around an embedded surface $\Sigma$,
are considered static models. A dynamical version was developed in \cite{AFS}, by embedding
the $4$-dimensional geometry in a $(4+1)$-dimensional Ricci-flat geometry. The discussion in
\cite{AFS} focuses on the case of the Taub-NUT $4$-dimensional gravitational instanton and
its embedding in a particular $5$-dimensional Ricci-flat geometry that can be viewed
as an evolving  Taub-NUT geometry, where the parameter of the Taub-NUT metric varies by
affine transformations, ensuring the Ricci-flat condition of the $5$-dimensional geometry. 
This $5$-dimensional geometry recovers the Sorkin solution of the Kaluza--Klein monopole
equations, \cite{Sorkin}.

\smallskip

More generally, the Campbell--Magaard embedding theorem shows that an
arbitrary analytic Riemannian manifold $M$ of dimension $\dim M=n$
can be locally embedded in a Ricci-flat Riemannian manifold of dimension $n+1$,
see \cite{Lidsey} for a discussion of applications, with particular attention to the
embeddings of $4$-dimensional (Riemannian) spacetimes in $(4+1)$-dimensional
Ricci flat manifolds. The result is extended in \cite{DaRom} to arbitrary signature
and to embeddings in Einstein $(n+1)$-dimensional manifolds. All these
results only work locally, and the embeddings need not  in general extend beyond 
a small neighborhood of an arbitrary chosen point. Any such embedding can be
viewed as a dynamical model for the static geometric model of matter described
by the $4$-dimensional geometry. In particular, the case of
interest to our setting is a $4$-dimensional geometry of the form 
$M\smallsetminus \Sigma$ as above, with an (orbifold) edge-cone
metric of cone angle $2\pi/\nu$, where a Ricci-flat $5$-dimensional embedding 
extends to a tubular neighborhood $\cN(\Sigma)\subset M\smallsetminus \Sigma$
with the edge-cone metric. The analyticity condition is needed for the local embedding.
Topological and
differentiable obstructions can 
impose constraints that limit the possibility of extending such embeddings beyond
the local existence. 

\smallskip

A particular form of the $5$-dimensional embedding of an Einstein $4$-dimensional
manifold $M$ with $R_{ij}=-\lambda g_{ij}$ is given in \cite{DaRom} \S III, with the $5$-dimensional 
metric of the form 
\begin{equation}\label{metric5dim}
 ds^2 = f(u) g_{ik} dx^i dx^k + \epsilon du^2 
\end{equation} 
with $\epsilon=\pm 1$ and with 
$$ f(u) =(\cosh((-\epsilon\Lambda/6)^{1/2} u)
+ (1-2\lambda/\Lambda)\sinh(((-\epsilon\Lambda/6)^{1/2} u))^2 , $$ 
depending on $\epsilon, \Lambda, \lambda$, where $\Lambda$ is the
cosmological constant of the Einstein metric on the $5$-dimensional geometry.
A similar geometric setting is considered in the context of brane world
collisions, \cite{GibbPope}.

\smallskip

One can consider, in addition to embeddings 
of the $4$-dimensional geometry $M$ into a $5$-dimensional 
product geometry $M\times I$, the possibility of topology changes,
namely of embeddings into topologically non-trivial $5$-dimensional
cobordisms. Physically acceptable conditions on the existence
of such topology changes are analyzed in \cite{DowGar}, in terms of
handlebody decompositions of the cobordism. Possibilities
include the case of pair production of Kaluza--Klein monopoles,
with the topology change from $S^3\times S^1$ to $S^4$ via a
$D^5\smallsetminus S^1\times D^4$, which admits Riemannian as well as 
causally continuous almost Lorentzian metrics, constructed using a Morse
function (see \cite{DowGar}). 

\smallskip
\subsection{Orbifolds as systems of quasi-particles}

A main reason for interpreting these types of $4$-dimensional orbifolds as geometric
models of quasi-particles is the presence of fractional quantum numbers. In the approach of \cite{AMS}
to geometric models of matter, the signature $\tau(M)$ %
is interpreted as a baryon
number, while the electric charge is determined by the self-intersection number of the surface 
at infinity. The Euler characteristic, on the other hand, does not play a direct role as a
quantum number in the geometric models of \cite{AMS},  %
unlike what typically happens 
in geometric models of the quantum Hall effect, where it is related to
the noncommutative Kubo formula for the transport coefficient 
(\cite{Bell}, \cite{CaHaMa}, \cite{MaMa1}, \cite{MaMa2}, \cite{MaMa3}, \cite{MaSe}).
In the more recent work of the first author and Nick Manton \cite{AtMan}, for models of matter based
on algebraic surfaces, baryon and lepton numbers are expressed in terms of both 
signature and Euler characteristic, with the signature measuring the difference between
the number of protons and the number of neutrons. The models considered in \cite{AMS} 
dealt with conformally self-dual manifolds. For these, the definitions agree with 
those of \cite{AtMan}.

\smallskip

Let $M$ be a $4$-dimensional compact orbifold $M$ with set of orbifold points given by
an embedded surface $\Sigma$, endowed with a self-dual (or anti-self-dual) orbifold 
edge-cone metric with cone angle $2\pi/\nu$ with $\nu\in \N$, $\nu\geq 2$. Let $W$ denote
the Weyl tensor, with $W^\pm$ the decomposition into self-dual and anti-self-dual part,
$E$ the traceless part of the Ricci tensor, and $R$ the scalar curvature.  In \cite{AtLe}
it is shown that one obtains an orbifold Euler characteristic and an orbifold signature,
respectively given by
\begin{equation}\label{chiorb}
\chi_{orb}(M)=\frac{1}{8\pi^2} \int_M \left( |W|^2 -\frac{1}{2} |E|^2 + \frac{1}{24} R^2 \right)\, dv(g) =
\chi(M) - (1-\frac{1}{\nu}) \chi(\Sigma) ,
\end{equation}
\begin{equation}\label{tauorb}
\tau_{orb}(M)= \frac{1}{12\pi^2} \int_M \left( |W^+|^2 - |W_-|^2 \right)\, dv(g) = \tau(M) -\frac{1}{3} (1-\frac{1}{\nu^2})[\Sigma]^2 ,
\end{equation}
where $[\Sigma]^2$ is the self-intersection number. 
As discussed in \cite{AtLe}, \cite{LoVia}, the orbifold Euler characteristic and signature of \eqref{chiorb}
and \eqref{tauorb} fit in the framework of the Kawasaki index theorem for orbifolds \cite{Kawa}.
Namely, as shown in \cite{LoVia}, 
there is an elliptic complex $\cK$, which depends on the orbifold $M$ and the edge-cone
metric, such that 
\begin{equation}\label{IndexK}
 {\rm Ind}(\cK) = \dim H^0 - \dim H^1 + \dim H^2 = \frac{1}{2}(15 \chi(M)- 29 \tau(M)) - 4 \chi(\Sigma)
+ 4 [\Sigma]^2. 
\end{equation}
The quantities $\chi_{orb}(M)$ and $\tau_{orb}(M)$ should be interpreted as
fractional quantum numbers for the orbifold $M$, viewed as modeling a system of
quasi-particles. 

\smallskip

The self-intersection number $[\Sigma]^2$ is the same as the Euler number of the
normal bundle of $\Sigma$ in $M$. In the complement $M_{reg}=M\smallsetminus \Sigma$,
the surface $\Sigma$ of orbifold points can be viewed, by analogy with the geometric models
of matter described in \cite{AMS}, as being the surface at infinity that contributes the electric
charge to the matter content. In order to take into account the orbifold structure properly, 
note that the normal bundle $\cN(\Sigma)$ of the inclusion of $\Sigma$ in $M$ is an orbifold
vector bundle, whose fibers are quotients $\R^2 / G_\nu$ where $G_\nu=\Z/\nu\Z$ is the
stabilizer of $\Sigma$. Thus, the self-intersection number should be replaced by the
orbifold Euler number of the normal bundle $\cN(\Sigma)$,
that is, $[\Sigma]^2_{orb}=\chi_{orb}(\cN(\Sigma))$. This is the 
rational valued Satake orbifold Euler characteristic \cite{Sat}, rather than the integer
valued orbifold Euler characteristic of good orbifolds $M=X/G$ considered
in \cite{AtSe} and \cite{HH}.
The fractional quantum number $[\Sigma]^2_{orb}=\chi_{orb}(\cN(\Sigma))$ represents 
the fractional electric charge of the system of quasi-particles. In ordinary matter, quarks
exhibit both fractional electric charge and fractional baryon numbers. Fractional baryon
numbers occur in quark-gluon plasma, in skyrmion models, and in models of baryogenesis
from scalar condensates. 

\smallskip

In the geometric models of matter developed in \cite{AFS}, \cite{AtMan}, \cite{AMS}
the interpretation of the signature as baryon number is compatible with viewing
baryon number as arising through a chiral symmetry breaking in QCD. The way
in which the chiral symmetry breaking is implicitly built into these models is through
an index theorem. Indeed, it is well known (see for instance \cite{Bakas2} and the survey \cite{Bakas})
that the axial anomaly is a topological density given by the Chern--Pontryagin class
and the integrated form of the anomalous axial current is expressed as an index
theorem of a Dirac operator. Thus, for instance, the fact that on Taub-NUT gravitational
instantons the relevant index is zero implies that chiral symmetry breaking is not
present, and this is consistent with the interpretation given in \cite{AFS} of the 
Taub-NUT as a geometric model for a particle with zero baryon number. 
In the setting we consider here the chiral symmetry breaking is still included in
the model in the form of an index theorem, but due to the presence of an orbifold structure,
the relevant form in which it appears is a Kawasaki index theorem for orbifolds, \cite{Kawa}.

\smallskip
\subsection{Composite systems}

Interpreting, as above, the $4$-dimensional orbifolds as quasi-particles allows for
interesting constructions of composite systems of quasi-particles, arising from 
natural geometric constructions of $4$-manifolds. However, as shown in \cite{AtLe},
\cite{LeBrun}, \cite{LoVia}, there are obstructions to the existence of Einstein
metrics on these composite systems, so it is not always possible to obtain such
systems as gravitational instanton models of matter. Some of the available
constructions that give rise to quasi-particle systems are connected sums
and branched coverings. We can regard these operations as ways of obtaining
composite systems, respectively, by a merging (fusion) and by a branching operation.
We describe these more in detail. 

\subsubsection{Connected sums}

As we recalled above, it is possible to endow with a self-dual metric
a connected sum of two self-dual $4$-manifolds \cite{DoFri}, 
under suitable conditions that can be identified in terms of twistor spaces.
The argument of \cite{DoFri} based on deformation theory was extended to
the case of self-dual $4$-dimensional orbifolds in \cite{LeBSing}, \cite{KoSing}, \cite{LoVia}.
In particular, an edge-cone metric is unobstructed if $H^2=0$ in the complex $\cK$ of \eqref{IndexK}.
If $(M_1,\Sigma_1)$ and $(M_2,\Sigma_2)$ are unobstructed self-dual
orbifolds, with set of orbifold points $\Sigma_i$ and 
with edge-cone metrics with the same cone angle $2\pi/\nu$, then the connected
sum $(M_1\# M_2, \Sigma_1\# \Sigma_2)$ can also be endowed with an unobstructed
self-dual orbifold-cone metric with the same cone angle and set of
orbifold points $\Sigma_1\# \Sigma_2$. This operation should be regarded as
a way of merging the systems of quasi-particles represented by the orbifolds
$(M_1,\Sigma_1)$ and $(M_2,\Sigma_2)$ into a combined system.

\subsubsection{Branched coverings}

One can view branched coverings as orbifold coverings. Thus, one can obtain $4$-dimensional
orbifolds $M$ where the set of orbifold points is given by an embedded (not necessarily connected)
surface $\Sigma$ by considering smooth $4$-manifolds that arise as branched coverings
of $M$ with branch locus $\Sigma$. In this type of construction, the branch locus $\Sigma$ is 
usually embedded in a way that can be highly knotted. Indeed,
a branched covering of order $n$ is determined by a representation of the fundamental
group $\pi_1(M\smallsetminus \Sigma)$ (which carries the information on the amount of 
knottedness of the embedding) in the symmetric group $S_n$.
It is known by the result of \cite{IoPier} that any orientable closed PL $4$-manifold is a $5$-fold simple branched covering of $S^4$, branched along an embedded surface $\Sigma$.
Moreover, it is known by \cite{Auroux} that every compact symplectic $4$-manifold is a
branched cover of $\C\P^2$ branched along a symplectic curve (immersed with cusps) in $\C\P^2$.
In \cite{AtBe} it was shown that the projective planes $\P^2(\C)$, $\P^2(\H)$ and $\P^2(\bO)$ are
branched coverings of $S^4$, $S^7$ and $S^{13}$, 
with branch locus respectively given by $\P^2(\R)$, $\P^2(\C)$ and $\P^2(\H)$. 
Passing from an orbifold geometry $(M,\Sigma)$ to an orbifold covering $(M',\Sigma')$
can be viewed as another way of obtaining composite systems of quasi-particles.

\subsubsection{Obstructions}

There are obstructions to be taken into account when forming composite systems
of quasi-particles with the methods described above. The first type of obstruction to
be taken into consideration arises in the deformation argument that ensures the existence
of self-dual metrics on the connected sum. For instance, there are no unobstructed
self-dual orbifold-cone metrics on $S^4$ with set of orbifold points an orientable
embedded surface $\Sigma$ of genus $g\geq 1$ (Corollary 1.9 of \cite{LoVia}),
while such unobstructed metrics exist for $\Sigma=S^2$ and $\Sigma=\R\P^2$, see \cite{AtLe},
and for any connected sum of an arbitrary number of $\R\P^2$, see \cite{LoVia}. 

\smallskip

If we insist on the requirement that the geometric models of matter should be
gravitational instantons, namely both self-dual and Einstein, then there are
also obstructions to the existence of Einstein metrics that one needs to
take into account. These obstructions can
be of topological nature or of differentiable nature. Topological obstructions
have been identified in \cite{AtLe}: the inequalities 
$$ 2 \chi(M)\pm 3\tau(M) \geq (1-\frac{1}{\nu}) ( 2\chi(\Sigma) \pm (1+\frac{1}{\nu}) [\Sigma]^2), $$
have to be satisfied for a $4$-dimensional orbifold $(M,\Sigma)$ to admit an 
Einstein edge-cone metric of cone angle $2\pi/\nu$. Differentiable obstructions
have been identified in \cite{LeBrun} using Seiberg--Witten theory: if $M$ admits
a symplectic form $\omega$ for which $\Sigma$ is a symplectic submanifold with
$(c_1(M)-(1-1/\nu)[\Sigma])\cdot [\omega]<0$, then for any $\ell\geq (c_1(M)-(1-1/\nu)[\Sigma])^2/3$
the pair $(M',\Sigma)$ with $M'=M\#^\ell \overline{\C\P^2}$ does not admit an Einstein edge-cone metric
(Theorem A of \cite{LeBrun}). 

\smallskip

For the reason described here above, the operation of connected sum that plays the role
of fusion rules giving rise to composite systems is not always well defined, in an unobstructed
way, within the class of gravitational instantons. Thus, it seems preferable, within the
context of obtaining sufficiently well behaved systems of quasi-particle, to relax the
assumptions that all the $4$-manifolds involved are gravitational instantons, and allow
for a larger class of  $4$-manifolds with edge-cone orbifold geometries.

\section{Braid groups, surface braids, and anyons}

In the previous section, we described geometric models of systems of quasi-particles
in terms of $4$-dimensional gravitational instantons $M$ with an orbifold structure and 
an edge-cone metric around an embedded surface $\Sigma$ of orbifold points. In this
section we show that these models exhibit interesting braiding structures that behave 
like physical anyons. It is well known that anyons arise only in two-dimensional systems,
due to the topology of configuration spaces that allows for interesting braid groups,
see \cite{Imbo}, \cite{Kitaev}, \cite{Lerda}. Indeed, anyons and representations of
braid groups have been considered in relation to quantum Hall systems, \cite{JSWW},
\cite{JGJJ}, \cite{MaSe}. While anyons do not arise in higher dimensions, we will see that
the presence of the $2$-dimensional surfaces $\Sigma$ of orbifold points allows
for the existence of non-trivial anyon states and interesting braid group representations
associated to the $4$-dimensional orbifold geometries that describe our quasi-particle systems.
We begin by analyzing the different forms of knottedness and braiding that are present
in our geometric setting and the role they play in describing properties of the
corresponding quasi-particle system. 

\subsection{Fundamental groups of surface complements}

An embedded $2$-dimensional surface $\Sigma$ in a $4$-dimensional
manifold $M$ has an associated fundamental group $\pi_1(M\smallsetminus \Sigma)$.
In the case of a $2$-knot, that is, an embedding $\iota: S^2\hookrightarrow S^4$, the
fundamental group $\pi_1 (S^4\smallsetminus \iota(S^2))$ plays a role analogous
to the knot groups $\pi_1(S^3\smallsetminus K)$ that measure the amount of knottedness
of embeddings $K$ of $S^1$ in $S^3$. 
Explicit presentations for fundamental groups $\pi_1(S^4\smallsetminus
\Sigma)$, with $\Sigma$ an embedded surface, are obtained in \cite{Kino}, by combining
the use of Wirtinger presentations of knot groups in $3$-dimensions and van Kampen's theorem.

\smallskip

In the case of $2$-dimensional surfaces $\Sigma$ embedded in smooth $4$-dimensional
manifolds, a phenomenon arises that is not present in the more familiar lower dimensional case,
namely the possibility of exotic knottedness: this refers to embedded surfaces that are topologically
but not smoothly isotopic, see \cite{Fin}, \cite{KimRub}. Since we are mostly interested here 
in discussing topological properties, and in particular representations of knot and braid groups 
arising from embeddings of surfaces in $4$-manifolds, we do not need to worry about the 
possible effect of exotic smoothness and knottedness. However, exotic smoothness can influence other
aspects of the geometry, such as conditions on the existence of appropriate metrics. 

\smallskip

Both the knot groups $\pi_1(S^3\smallsetminus K)$ and the fundamental
groups $\pi_1(S^4\smallsetminus \Sigma)=\pi_1(\R^4\smallsetminus \Sigma)$, 
with $\Sigma$ an embedded surface, are examples of a larger class of groups,
called $C$-groups, see \cite{Kulikov}. These are defined by the existence of a presentation of
the form $\langle x_1, \ldots, x_n \,|\, R_\alpha \rangle$ where the relations
$R_\alpha$ with $\alpha=(\alpha_i)_{i=1,2,3}$ are all conjugations of the form 
$R_\alpha = x_{\alpha_1} x_{\alpha_2} x_{\alpha_1}^{-1} x_{\alpha_3}^{-1}$.
If $\Sigma$ is a smoothly embedded $2$-dimensional orientable compact 
surface in $S^4$, then $\pi_1(S^4\smallsetminus \Sigma)$ is a $C$-group
whose abelianization is $\pi_1(S^4\smallsetminus \Sigma)^{ab}=\Z^k$, where
$k$ is the number of components of $\Sigma$, see Theorem 1 of \cite{Kulikov}. 
It is shown in \cite{Kulikov} that any $C$-group can be realized as 
$\pi_1(S^4\smallsetminus \Sigma)$, with $\Sigma$ an embedded surface,
though not all $C$-groups are fundamental groups of complements of
$2$-knots $S^2 \hookrightarrow S^4$. However, as we pointed out before, 
the complements $S^4\smallsetminus \Sigma$ with $g(\Sigma)\geq 1$ do 
not carry unobstructed self-dual metrics, by \cite{LoVia}, so not all these possibilities
will arise from geometric models of quasi-particle systems with good composition
properties. In any case, the general result of \cite{Kulikov} shows that the class of 
fundamental groups $\pi_1(M \smallsetminus \Sigma)$ can be highly nontrivial.

\smallskip

However, there are also many significant examples  where
the fundamental group $\pi_1(M\smallsetminus \Sigma)$
does not carry enough interesting information. For instance, in the case of an embedding of
an algebraic curve $C$ in the plane $\C^2$, if the curve is smooth, then the fundamental
group $\pi_1(\C^2\smallsetminus C)$ is cyclic generated by a loop transverse to the curve.
It was shown in \cite{Deligne} \cite{Ful} that even in the case of nodal singularities the fundamental
group of the complement is abelian. This fact suggests that, while generally interesting, the
groups $\pi_1(M\smallsetminus \Sigma)$ are not the correct fundamental groups to consider in 
our setting, to obtain a physical system with interesting fractional statistics.
  
\smallskip

For more general singular plane curves, however, the fundamental group of the
complement is not necessarily abelian and can be very interesting. A significant 
example was already described by Zariski in \cite{Zariski} (see also \cite{Zariski2}):
in the case where $C$ is a sextic with six cusps, if the cusps lie on a conic,
the fundamental group of the complement is $\Z/2\Z \star \Z/3\Z$, while 
if the six cusps do not lie on a conic the fundamental group is $\Z/2\Z \times \Z/3\Z$.
A cohomological interpretation of the Zariski example is discussed, for instance, 
in \S 4 of \cite{Dimca}.

\smallskip

Moreover, for singular curves there is an interesting relation between the
fundamental group $\pi_1(\C^2\smallsetminus C)$ and braid 
groups, in the form of {\em braid monodromy}. 
Unlike the case of a smooth or nodal algebraic curve $C$ in $\C^2$,
for a more general planar curve with arbitrary singularities, the fundamental
group $\pi_1(\C^2\smallsetminus C)$ has a more complicated structure. 
For the general case of singular algebraic curves $C$ in $\C^2$, a presentation of
$\pi_1(\C^2\smallsetminus C)$ was constructed in \cite{Libgober}, with
the property that the $2$-dimensional CW complex associated to the
group presentation, with one $0$-cell,
a $1$-cell for each generator and a $2$-cell for each relation, 
is homotopy equivalent to the complement $\C^2\smallsetminus C$.
The presentation is based on the braid monodromy construction of \cite{Mois}.
A linear projection of $\C^2$ onto a line $L=\C$ determines a locally trivial
bundle $\C^2\smallsetminus C \to L$ and a homomorphism (the braid
monodromy) of $\pi_1(\C^2\smallsetminus C)$ to the group of diffeomorphisms
of the fiber that fix the intersection with $C$, identified with a braid group,
see \cite{Libgober}, \cite{Mois}.
We will see below that a similar structure with braid group representations
will provide the source of anyons in our models. 

\subsubsection{Orbifold fundamental group}

In general if $M$ is a good orbifold, with singular locus $M_{sing}=\Sigma$ of real codimension two,
the orbifold fundamental group $\pi_1^{orb}(M)$ is given (see \cite{Thur}, \S 13) by the quotient
\begin{equation}\label{orbpi1}
\pi_1^{orb}(M) = \pi_1(M_{reg}) / H,
\end{equation}
where $M_{reg}=M \smallsetminus M_{sing} =M \smallsetminus \Sigma$ is the set of 
regular points of the orbifold and
$H$ is the normal subgroup generated by the classes $\gamma_j^{\nu_j}$ in $\pi_1(M_{reg})$,
where the $\gamma_j$ are loops encircling the connected components $M_{sing,j}=\Sigma_j$ of $M_{sing}=\Sigma$, and $\nu_j \in \N$ is the order of the stabilizer $G_j$ of the component $M_{sing,j}$.
Thus, in the kind of geometry that we consider here, the fundamental group $\pi_1(M\smallsetminus \Sigma)$
considered above, should be replaced by the orbifold fundamental group 
$\pi_1^{orb}(M) =\pi_1(M\smallsetminus \Sigma)/H$. 

\smallskip

One can view the sphere 
$S^4$ as a quotient of $\C\P^2$ by $\Z/2\Z$ with branch locus $\R\P^2$ (see \cite{AtBe}
for similar descriptions of $S^7$ and $S^{13}$). On the complement 
$M\smallsetminus \Sigma = S^4\smallsetminus \R\P^2$ 
consider the Hitchin family of self-dual Einstein orbifold
edge-cone metrics with cone angle $2\pi/(k-2)$,
constructed in \cite{Hitchin}. The fundamental group is 
$\pi_1(S^4\smallsetminus \R\P^2)=\Z/2\Z$.  When $k$
is even $\pi_1^{orb}(M)=\Z/2\Z$, while it is trivlal when $k$ is odd. 
The same holds for composite systems obtained by connected sums,
with $M_{reg}=S^4 \smallsetminus \#^\ell \R\P^2$, with $\pi_1(M_{reg})=\Z/2\Z$.  
As another example, consider the Atiyah-LeBrun edge-cone metrics defined in \cite{AtLe} p.21, 
on $M_{reg}=S^4\smallsetminus S^2$,
with the standard unknotted embedding of $S^2$, with cone angle $2\pi/\nu$. The associated 
orbifold fundamental group is $\pi_1^{orb}(M)=\Z/\nu\Z$.  

\smallskip

It is clear from these examples that,
in such cases of interesting geometric models of matter with orbifold structure, the orbifold
fundamental group is too simple to give rise to any interesting braiding that can be
interpreted in terms of anyon states. We will review briefly the relation between braid groups,
fractional statistics and anyons, and then we will revisit the geometry of the embedded
surfaces $\Sigma=M_{sing}$ in order to identify the correct source of interesting braiding.

\subsection{Configuration spaces and braid groups}
Let $X$ be a smooth manifold. Consider the space $F_n(X)=X^n \smallsetminus \Delta$,
that is, the complement of the diagonals in $X^n$, 
$$ F_n(X) =\{ (x_1,\ldots, x_n) \in X^n\,|\, x_i \neq x_j, \, \forall i\neq j, \, i,j=1,\ldots, n \}. $$
The symmetric group $S_n$ acts freely on $F_n(X)$. The configuration spaces of $X$ are the
quotients
\begin{equation}\label{ConfnX}
\Conf_n(X) := F_n(X)/S_n .
\end{equation}
Configuration spaces play an important role in physical models, where
they describe distinct point particles and provide a geometric setting for
quantum mechanical spin-statistics results, see \cite{At},  \cite{AtBel}, \cite{AtSu}, \cite{AtSu2}.
The Pauli sign for fermions was interpreted in \cite{Berry}, using Schwinger representations of spin, 
as a topological phase arising from noncontractible loops in a nonorientable configuration space.
In the context of geometric models of matter, the $n=2$ configuration space plays an
important role in the construction of \cite{AtMan2}. Configuration spaces also 
provide the background for the formulation of the multi-Taub-NUT geometries
considered in \cite{AFS}, which can be regarded as a geometric  approach towards 
some of the ideas in Feynman's thesis \cite{Feynman}. 

\smallskip

The braid groups of $X$ are the fundamental groups
\begin{equation}\label{BnX}
B_n(X) := \pi_1(\Conf_n(X)).
\end{equation}
The quotient description  \eqref{ConfnX} implies that
the fundamental groups are related by a sequence 
\begin{equation}\label{pi1Conf}
 1\to \pi_1(F_n(X)) \to \pi_1 (\Conf_n(X)) \to S_n \to 1 ,
\end{equation} 
while the higher homotopy groups satisfy $\pi_i(F_n(X)) = \pi_i (\Conf_n(X))$ for $i\geq 2$.

If $X$ is a manifold of dimension $m=\dim X > 2$, then it is known 
(see for instance \cite{KalSa} and \cite{Imbo}) that
$\pi_1(F_n(X))=\pi_1(X)^n$ and that the braid group is a wreath
product 
\begin{equation}\label{wreathBn}
B_n(X) = \pi_1(X)\wr S_n=\pi_1(X)^n \rtimes S_n ,
\end{equation}
while the higher homotopy groups satisfy
\begin{equation}\label{pige2X}
\pi_i(\Conf_n(X)) = \pi_1(X)^n, \ \ \ \text{ for } 2\leq i\leq m-2 ,
\end{equation}
see Theorem 1.2 of \cite{KalSa}.

\smallskip

In particular, we are interested in the case where $X=M\smallsetminus \Sigma$
where $M$ is a smooth compact $4$-manifold and $\Sigma$ is a smoothly embedded 
compact $2$-dimensional surface (not necessarily connected). In this case we have
\begin{equation}\label{BnMS}
B_n(M\smallsetminus \Sigma)= \pi_1(M\smallsetminus \Sigma) \wr S_n,
\end{equation}
A presentation of $\pi_1(M\smallsetminus \Sigma)$ can then be used to
obtain an explicit presentation for the braid groups $B_n(M\smallsetminus \Sigma)$. 
For the second homotopy group we simply have
\begin{equation}\label{pi2MS}
\pi_2(\Conf_n(M\smallsetminus \Sigma))= 
\pi_2 (M\smallsetminus \Sigma)^n .
\end{equation}

\smallskip

For example, if $\iota: S^2 \hookrightarrow S^4$ is a smoothly embedded 2-knot in 
$S^4$, such that $\pi_1(S^4\smallsetminus \iota(S^2))=\Z$, then
by \cite{Levine} the complement $S^4\smallsetminus \iota(S^2)$
has the homotopy type of $S^1$ and we obtain 
$B_n(S^4\smallsetminus \iota(S^2)) = \Z \wr S_n$ and
$\pi_2(\Conf_n(S^4\smallsetminus \iota(S^2)))=0$.

\smallskip

The fact that the braid groups for manifolds of dimension at least three
have a simple structure as wreath products of the fundamental group of
the manifold and the symmetric group means that, unlike the case of
dimension two, there are no anyon states arising from representations
of the braid groups of the ambient space. Indeed, 
for a system of $n$ identical particles on a smooth manifold $X$, with configuration space
$\Conf_n(X)$, the set of irreducible unitary representations of the braid group 
$B_n(X)=\pi_1(\Conf_n(X))$ labels inequivalent quantizations of the classical system.
These can exhibit different possible statistics, which include bosons and fermions, as
well as parastatistics, generalized parastatistics, and anyons. 
Parastatistics arise from higher dimensional representations of the symmetric groups,
while fermions and bosons correspond to $1$-dimensional representations. 
In the case of a simply connected manifold $X$ of dimension $\dim X\geq 3$, 
the braid groups are just symmetric groups by \eqref{wreathBn}, hence one can
only obtain fermions and bosons, or parastatistics. In particular, the only 
$1$-dimensional (scalar) quantizations are either fermions or bosons. In cases of manifolds with
$\dim X\geq 3$ with non-trivial fundamental groups, one obtains generalized parastatistics 
(see \cite{Imbo}).  In the case of $2$-dimensional manifolds, however, the situation is more
interesting. On a $2$-dimensional surface the braid group $B_n(X)$ is not simply a 
wreath product as in \eqref{wreathBn}, but has a more interesting
structure computed in \cite{Bir1}, \cite{Bir2}. 
This allows for more general exotic statistics, where even in the scalar case one
can have statistics that are not fermions or bosons, but more general anyons,
depending on an angle $\theta$. Non-abelian anyons arise from higher dimensional
representations of the braid groups of $2$-dimensional manifolds. 

\smallskip
\subsubsection{Orbifold braid groups}

In the case of a $4$-dimensional orbifold geometry $(M,\Sigma)$ with $M_{reg}=M\smallsetminus \Sigma$
and $2$-dimensional $M_{sing}=\Sigma$, one can replace the braid groups 
$B_n(M_{reg})=\pi_1(M_{reg})\wr S_n$ with the orbifold braid group as in \cite{MaSe},
\begin{equation}\label{orbBn}
B^{orb}_n(M) = \pi_1^{orb}(M)\wr S_n = \pi_1^{orb}(M)^n \rtimes S_n,
\end{equation}
with $\pi_1^{orb}(M)$ as in \eqref{orbpi1}. 

\smallskip

For example, in the case of the orbifold structure on $M_{reg}=S^4\smallsetminus \R\P^2$ given
by the Hitchin metrics of \cite{Hitchin} with cone angle $2\pi/(k-2)$, the orbifold braid
groups are $B_n^{orb}(M)=S_n$ for $k$ even and $B_n^{orb}(M)=\Z/2\Z\wr S_n$ for $k$ odd, while
for the Atiyah-LeBrun orbifold structures on $M_{reg}=S^4\smallsetminus S^2$ with cone angle
$2\pi/\nu$ the orbifold braid groups are $B_n^{orb}(M)=\Z/\nu\Z\wr S_n$. 

\smallskip

Representations of orbifold 
braid groups of the $4$-dimensional orbifold geometries $(M,\Sigma)$
can determine parastatistics and generalized parastatistics.
Anyon representations associated to orbifold braid groups of $2$-dimensional
orbifolds were classified in \cite{MaSe}, in terms of orbifold line bundles and
Seifert invariants, in the context of quantum Hall models.
Since anyons only arise from $2$-dimensional geometries, the natural source of
anyons in our models are the surfaces $\Sigma$ of orbifold points and the
braid representations arising from associated surface braids. We explain this
in the rest of this section.

\smallskip
\subsection{Surface braids}\label{2braid}

Surface braids are a two-dimensional generalization of braids, initially
introduced by Oleg Viro and developed by Kamada, \cite{Kama}, \cite{Kama2}.
A surface $m$-braid is a smooth $2$-dimensional surface $S$, smoothly embedded
in $D^2 \times D^2$, such that the second projection $P_2: D^2 \times D^2 \to D^2$
restricted to $S$ is an $m$-fold branched cover $P: S\to D^2$. The preimage
$P_2^{-1}(\partial D^2)\cap S \subset D^2 \times S^1$ is a closed ordinary $m$-braid $\beta$. 
(Note: the terminology ``surface braid" we use here is often used in the literature for
the more restricted case where $\beta$ is the trivial braid, with ``braided surface" used
for this more general case, \cite{Kama2}.)

\smallskip

Let $b(S)\subset D^2$ denote the set of branch points of the $m$-fold branched covering
map $P: S\to D^2$. 
Let $\gamma(t)$ be a path in  $D^2 \smallsetminus b(S)$
that represents a class in the fundamental group $\pi_1(D^2 \smallsetminus b(S))$, 
computed, for example, with respect to a base point on the boundary $\partial D^2$.
Taking 
\begin{equation}\label{rhoSgamma}
\rho_S(\gamma)(t):=P_1(S \cap P_2^{-1}(\gamma(t))), 
\end{equation}
where $P_i: D^2 \times D^2 \to D^2$
are the two projections, determines a path in $\Conf_m(D^2)$. This
determines the {\em braid representation} 
\begin{equation}\label{braidrepS}
\rho_S : \pi_1(D^2 \smallsetminus b(S)) \to \pi_1(\Conf_m(D^2)) =B_m(D^2).
\end{equation}

\smallskip

A closed surface braid is similarly defined in \cite{Kama}, \cite{Kama2} as a smoothly
embedded $S$ in $D^2 \times S^2$, such that the restriction to $S$ of the
projection to $S^2$ is an $m$-fold branched covering map $P:S \to S^2$. 
In our setting, we consider a more general form of closed surface braid, where
$S^2$ is replaced by an arbitrary compact smooth $2$-dimensional $\Sigma$.
In particular, we consider the setting as above, where $\Sigma$ is a smoothly
embedded smooth compact $2$-dimensional manifold in a compact
$4$-dimensional manifold $M$.  If $\Sigma$ has several connected
components, we focus on the neighborhood of only one component. 
Let $\cN(\Sigma)$ be a tubular neighborhood of $\Sigma$ in $M$.
Locally, over an open ball $D^2\subset \Sigma$, the tubular neighborhood 
is isomorphic to a product $D^2 \times D^2$. By indentifying $\cN(\Sigma)$
with the unit disc bundle of the normal bundle of the embedding 
$\Sigma \hookrightarrow M$, we write $P_\cN: \cN(\Sigma) \to \Sigma$ for the
corresponding projection with fiber $D^2$. We define a closed surface braid in $M$
as an embedded surface $S$ in $\cN(\Sigma)$ such that the restriction to $S$
of the projection $P_\cN: \cN(\Sigma) \to \Sigma$ is a $m$-fold branched
cover $P: S \to \Sigma$. In the case where $\Sigma$ is an unknotted $S^2$
this recovers the original formulation of closed surface braids of Viro and Kamada. 

\smallskip 

Let $a_1, \ldots, a_g, b_1,\ldots, b_g$ be a set of generators for $\pi_1(\Sigma)$,
where $g=g(\Sigma)$ is the genus. Consider a choice of representatives 
$a_i(t), b_i(t)$ given by paths in $\Sigma$. Let $P_\cN: \cN(\Sigma) \to \Sigma$
be the projection as above and let $S \subset \cN(\Sigma)$ be a surface $m$-braid.
Then $P_\cN^{-1}(a_i)\cap S$ and $P_\cN^{-1}(b_i)\cap S$ are closed ordinary $m$-braids
$\beta_{a_i}, \beta_{b_i}$ in $D^2 \times a_i=D^2\times S^1$ and $D^2 \times b_i 
=D^2 \times S^1$, respectively. 

\smallskip

Given a closed surface $m$-braid  $S$ in $\cN(\Sigma)$, which is an $m$-fold branched 
cover of $\Sigma$ branched along a set of points $b(S)\subset \Sigma$, the associated 
braid representation is given by the group homomorphism
\begin{equation}\label{braidrepSc}
\rho_S: \pi_1(\Sigma\smallsetminus b(S)) \to \pi_1(\Conf_m(D^2))=B_m(D^2), 
\end{equation}
obtained as above by setting $\rho_S(\gamma)=P_1(S\cap P_2^{-1}(\gamma))$,
with $P_2$ the projection of the bundle $\cN(\Sigma)\to \Sigma$ and $P_1$ the
local projection in the fiber direction at a point in $S\cap P_2^{-1}(\gamma)$.

\smallskip
\subsection{Orbifold normal bundle} 

In the case of the geometries we are considering, the embedded surface $\Sigma$
in the $4$-manifold $M$ is the set of the orbifold points $M_{sing}=\Sigma$ of $M$.
Thus, the normal bundle $\cN(\Sigma)$ is in fact an orbifold bundle. We assume that
$M$ is a good orbifold, covered by a compact smooth $4$-dimensional
manifold $X$ with an action of a finite group $G$, so that $M=X/G$. For simplicity, we
can assume that the set of orbifold points is a connected surface and that $G=\Z/\nu\Z$.
Then the orbifold
bundle $\cN(\Sigma)$ is orbifold covered by the normal bundle $\cN(\tilde\Sigma)$ 
of $\tilde\Sigma$, the preimage of $\Sigma$ in $X$. 
A section $\sigma$ of $\cN(\Sigma)$ in general position intersects the zero section
in a finite set of points $Q$. The preimage in $\cN(\tilde\Sigma)$ then determines
a $\nu$-fold covering $S$ of $\Sigma$ branched at $Q$. We can identify $S$ with
a surface braid, a $\nu$-fold cover of $\Sigma$ branched over $Q=b(S)$.

\smallskip

We obtain in this way, from an orbifold $(M,\Sigma)$ with cone angle $2\pi\nu$, 
surface braids $S$ that are $\nu$-fold branched covers $P: S \to \Sigma$. In particular,
if we have a fixed geometry $(M,\Sigma)$ that admits a family of orbifold edge-cone 
metrics with cone angles $2\pi/\nu$ for any $\nu\in \N$, $\nu\geq 2$, we obtain
surface $\nu$-braids $S$ for all $\nu\in \N$, $\nu\geq 2$, in the respective lifts $\cN(\tilde\Sigma)$
of the orbifold normal bundle $\cN(\Sigma)$. Each of these surface braids determines a
braid representation $\rho_S: \pi_1(\Sigma\smallsetminus b(S)) \to \pi_1(\Conf_\nu(D^2))=B_\nu(D^2)$.

\smallskip

In addition to considering sections of the orbifold normal bundle $\cN(\Sigma)$, we can
also consider multisections, 
given in local orbifold charts as $\Z/\nu\Z$-equivariant maps to $S^n(F)=F^n/S_n$, the
symmetric product of the fiber $F\simeq D^2$ of the unit normal bundle $\cN(\tilde\Sigma)$ that 
orbifold covers $\cN(\Sigma)$. Any such multisection $S$ determines an $\ell$-fold branched
cover of $\tilde\Sigma$ branched where the multisection meets the diagonals in the symmetric
product, hence an $\ell \nu$-fold branched cover of $\Sigma$, whose branch locus we again
denote by $b(S)$. This gives an associated braid representation 
$\rho_S: \pi_1(\Sigma\smallsetminus b(S)) \to \pi_1(\Conf_{\ell\nu}(D^2))=B_{\ell\nu}(D^2)$.
Thus, given an edge-cone metric with cone angle $2\pi/\nu$, by considering all multisections
of the orbifold normal bundle to $\Sigma$ that are in general position, we obtain 
braid representations in all the braid groups $B_n(D^2)$ with $n=\ell\nu$ for some $\ell\in\N$. 

\smallskip
\subsection{Anyons and vortices}

In the description above, anyon states arise from surface braids given by
multisections $S$ of the orbifold normal bundle of the surface $\Sigma$ of 
orbifold points in $M$. As such, these behave like extended objects. However,
a localization to pointlike objects is taking place, through the fact that the 
associated braid representation depends on the branch points $b(S)$
and the structure of branched cover $S\to \Sigma$, which is described 
by local monodromy data. 

\smallskip

Given the data of the surface with marked points $(\Sigma, b(S))$, one can
consider associated vortex moduli spaces, in the form of symmetric products
${\rm Sym}^n(\Sigma,b(S))$, see \cite{GaPra} and \S 5.7 of \cite{MOY}.
Note that the fundamental group of the symmetric products ${\rm Sym}^n(\Sigma, b(S))$
for $n>1$ are simply given by the abelianization of the fundamental group of
the surface with marked points, see Remark 5.8 of \cite{SGA1} and Lemma 2.3
of \cite{MaSe}. However, the associated configuration spaces ${\rm Conf}_n(\Sigma\smallsetminus b(S))$
determine braid groups $B_n(\Sigma,b(S)):=\pi_1({\rm Conf}_n(\Sigma\smallsetminus b(S)))$,
with explicit Artin presentations as in \cite{Bir1}, \cite{Bir2}. Thus, one can view the anyon
states described above, arising from the braid representation $\rho_S: \pi_1(S\smallsetminus b(S))
\to B_{\ell\nu}(D^2)$ as the first level of a more general construction that involves
also braid representations where $\pi_1(S\smallsetminus b(S))$ is replaced by the $B_n(\Sigma,b(S))$
for higher $n>1$. For the purpose of the present paper we focus only on the anyon
states associated to the $n=1$ level and the braid representation $\rho_S: \pi_1(S\smallsetminus b(S))
\to B_{\ell\nu}(D^2)$, which suffice, as we show in the next section, to obtain representations
that are universal for quantum computing. 

\smallskip
\section{Orbifold edge-cones as a quantum computer}

An important question regarding anyon systems is whether
the associated braid representations are universal for
quantum computing, which means unitary representations
that span densely the group $SU(2^N)$ of quantum gates 
for a system of $N$-qbits.  In this section we address this
question for the anyon systems constructed in the previous
section and we show that, in one of the simplest cases of data
$(M,\Sigma)$ given by the Atiyah--LeBrun orbifold edge-cone
metrics on $S^4\smallsetminus S^2$ with cone angle $2\pi/\nu$
one does indeed obtain a braid representation that is
universal for quantum computing. 

\smallskip

To put this question in context, note how  %
intriguing connections between spacetime geometry and quantum computation
have emerged recently in theoretical physics. The possibility of spacetime
being emergent from quantum information and entanglement via quantum
error correcting codes and tensor networks was proposed 
in the context of AdS/CFT correspondences, see  \cite{Happy}. The idea of
the universe itself as a quantum computer was discussed in \cite{Lloyd}.

\smallskip

Here we take a different viewpoint and we suggest that the $4$-dimensional
orbifolds that arise as geometric models of matter allow for the presence of
certain anyon representations that behave like a topological quantum computer,
and we show that very simple examples can be constructed for which the resulting 
quantum computer is universal. 

\smallskip
\subsection{Braided surfaces and universal quantum computers}

One of the main questions regarding physical systems that exhibit anyon statistics
is whether they can determine unitary representations of the relevant braid groups
that span densely the group $SU(2^N)$, for a system of $N$-qbits. 
This property ensures that arbitrary quantum circuits can be
approximated with a controllable error by elements in the representation, that is,
that the representation determines a universal quantum computer, \cite{Kitaev2}.
The Fibonacci anyons are an example of an anyon system satisfying this 
universality property, \cite{Tre}.

\smallskip

For a disc $D^2$, the braid group $B_n=B_n(D^2)$ is given by the Artin presentation
$$ B_n = \langle \sigma_1,\ldots, \sigma_{n-1}\,|\, \sigma_i \sigma_{i+1} \sigma_i = \sigma_{i+1} \sigma_i \sigma_{i+1}  \text{ and }  
\sigma_i\sigma_j=\sigma_j\sigma_i \text{ for } |i-j|\geq 2 
\rangle. $$
A class of representations of the braid groups with very useful applications to quantum
computation is given by the Jones representations of \cite{Jones}. These are obtained
by mapping, using the Kauffman bracket, the braid group algebra $\F[B_n]$ 
to the Temperley-Lieb algebra $TL_n(A)$, which is a quotient of the Hecke
algebra $H_n(q)$, for $q=A^{-4}$. The algebra $TL_n(A)$ is then identified 
with a sum of matrix algebras $M_{n_i}(\F)$, where $\F$ is the field of rational functions in the
variable $A$, and restriction to these building blocks defines the Jones representations
of the braid group, see the recent survey \cite{Del} for a quick overview. 
A construction in terms of braid groups of the Hecke algebra representations
associated to the one-variable Jones polynomial was given in \cite{Law}.

\smallskip

A crucial result in topological quantum computing is the fact that certain Jones
unitary representations of the braid groups $B_n=B_n(D^2)$ 
determine a universal quantum computer, \cite{Freedman}, \cite{Freedman2}.

\smallskip

In particular, the results of \cite{Freedman}, \cite{Freedman2}, showing that 
Jones representations can give rise to a universal quantum computer, can be used to
show that the Jones polynomial can be approximated efficiently by a quantum 
computer, see \cite{Del} and also \cite{FKW}, \cite{SchJo}. A new approach to the
Jones polynomial, currently being developed in \cite{AtJor}, presents a
different possible approach to questions about its efficient computability,
in the classical and quantum setting. 

\smallskip

A  property of quantum computation, which is very useful in order to obtain this 
type of density results, is the fact that the $1$-qbit gates given by elements of
$SU(2)$ together with the CNOT gate 
$$ \begin{pmatrix} 1 & 0 & 0 & 0 \\ 0 & 1 & 0 & 0 \\ 0 & 0 & 0 & 1 \\ 0 & 0 & 1 & 0 \end{pmatrix} $$
form a universal set for quantum computation, so that arbitrary gates in
$SU(2^N)$ can always be decomposed as tensor products of CNOTs and
$1$-qbit gates. Thus, in order to prove that certain unitary representations
$\rho: B_n \to U(\cH)$ of braid groups are universal for quantum computing,
it suffices to show that they approximate with arbitrary precision all the 
$1$-qbit gates and CNOT.

\smallskip

For example, in \cite{Freedman2} the state space $(\C^2)^{\otimes \ell}$ of
$\ell$-qbits is embedded in the space $V(D^2,3\ell)$ assigned by a TQFT to
the disc $D^2$ with $3\ell$ marked points, so that the action of $B_{3\ell}$ on
$V(D^2,3\ell)$ is intertwined, via the embedding, with unitary operators
acting on the state space $(\C^2)^{\otimes \ell}$. The TQFT considered
in \cite{Freedman2} is a Chern--Simons theory at a $5$-th root of unity. 
The action of $B_3$ on $V(D^2,3)=\C^2$ gives the $1$-qbit gates, while it is
possible to obtain a desired $2$-qbit gate by an approximation algorithm
(Theorem 2.1 of \cite{Freedman2}). This result is reformulated in \cite{Freedman}
in terms of the Jones representations at $q=e^{\pm 2\pi i/5}$. Note that in 
Chern--Simons theory level $\ell$ corresponds to $q^{2\pi i/(\ell+2)}$, \cite{Witten}.
This realization via Chern--Simons at a $5$-th root of unity
of a braid group representation that is universal for quantum
computing suffices for our purposes. It allows us to identify sufficiently simple
examples of geometric models of matter $(M,\Sigma)$ with orbifold geometry, 
where the braid representations of the surface braids determined by 
multisections of the orbifold normal bundle to $\Sigma$ admit unitary representations
that are universal for quantum computation.

\smallskip

Consider the example of the Atiyah--LeBrun orbifold edge-cone metric 
on $M_{reg}=S^4 \smallsetminus S^2$ with cone angle $2\pi/\nu$, as in \cite{AtLe}.
The embedding of $\Sigma=S^2$ in $M=S^4$ is standard unknotted and the normal bundle 
$\cN(\tilde\Sigma)$ can be identified with $S^2\times D^2$ with $\Z/\nu\Z$
acting on the fiber $D^2$, and with $\cN(S^2)$ the orbifold quotient. We use the notation
$D^2_f$ and $D^2_b$ to distinguish between the $2$-disc $D^2_f$ in the fiber of $\cN(S^2)$
and the choice of a $2$-disc $D^2_b$ in the base $S^2$. An $\ell$-multisection
is then an $\ell \nu$-fold branched cover $S$ of $S^2$ in $S^2\times D_f^2$, that
is, a closed surface $\ell \nu$-braid in the sense of \cite{Kama}, \cite{Kama2}. 

\smallskip

Given a closed surface $\ell \nu$-braid as above, choose one of the branch
points $x_0$ in $S^2$ and a disc $D_b^2\subset S^2$ that is the complement of a
small neighborhood of the chosen branch point. The restriction of the branch
cover projection $S\to S^2$ of the closed surface braid to this disc determines
a braided surface $\hat S$ in $D_b^2\times D_f^2$ that is an $\ell \nu$-fold branched 
cover of $D_b^2$ branched at $b(\hat S)=b(S)\smallsetminus \{ x_0 \}$. The intersection 
$\hat S\cap \partial D_b^2 \times D_f^2$ is the closure of a braid $\beta \in B_{\ell\nu}$,
which is in general nontrivial. Let $n=\# b(\hat S)$ and let $\gamma_1, \ldots, \gamma_n$
be a set of generators (a Hurwitz arc system) of the fundamental group 
$\pi_1(D_b^2 \smallsetminus b(\hat S))$, computed with respect to a chosen 
basepoint on the boundary $\partial D_b^2$. The braid representation 
$\rho_{\hat S}: \pi_1(D_b^2 \smallsetminus b(\hat S)) \to B_{\ell\nu}(D_f^2)$ is then 
determined by the images $\beta_k =\rho_{\hat S}(\gamma_k)$, for $k=1,\ldots,n$.
The element $(\beta_1,\ldots,\beta_n)\in  B_{\ell\nu}(D^2)^n$ is the braid system
of the braided surface $\hat S$. 

\smallskip

For a braided surface $\hat S$ that is an $\ell\nu$-fold branched cover of $D^2$,
one can give a characterization of all braid systems 
with the property that $\hat S\cap \partial D_b^2 \times D_f^2$ is a given closed braid 
$\beta \in B_{\ell\nu}(D_f^2)$. 
Such braid systems are given by all the $n$-tuples $(\beta_1,\ldots,\beta_n)\in  B_{\ell\nu}(D^2)^n$ 
with the property that each $\beta_k$ is a conjugate of a standard generator $\sigma_i$ of 
$B_{\ell\nu}(D^2)^n$ or an inverse $\sigma_i^{-1}$ and with $\beta = \beta_1\cdots \beta_n \in B_{\ell\nu}(D_f^2)$. 
The Hurwitz action of the braid group $B_n=B_n(D^2)$ on the $n$-fold 
product $B_{\ell\nu}(D^2)^n$ is given by 
$$ \sigma_i : (\beta_1,\ldots,\beta_i, \beta_{i+1},\ldots\beta_n)\mapsto 
(\beta_1,\ldots,\beta_{i-1}, \beta_i \beta_{i+1} \beta_i^{-1}, \beta_i,\beta_{i+2},\ldots\beta_n). $$
Elements $(\beta_1,\ldots,\beta_n)$ that are in the same orbit of the Hurwitz action
correspond to equivalent braided surfaces, that is, braided surfaces 
related by a fiber preserving diffeomorphism of
$D_b^2\times D_f^2$ relative to the boundary $\partial D_b^2 \times D_f^2$,
see \cite{Kama2}. In particular, for $n=\ell\nu-1$, one can consider the standard braided 
surface of degree $\ell\nu$ with $\ell\nu-1$ branch points, which corresponds to the
$n$-tuple of the standard generators $(\sigma_1,\ldots,\sigma_n)\in B_{n+1}^n$. 
There are in this case $(n+1)^{n-1}$ elements in the Hurwitz orbit, \cite{Hump}. 

\smallskip

This example shows that, for each $\nu$ and $\ell$, there is a particular choice
of a braided surface of degree $\ell\nu$ with $\ell\nu-1$ branch points whose
braid system is the standard set of generators $(\sigma_1,\ldots,\sigma_{\ell \nu -1})$
of the braid group $B_{\ell\nu}(D^2)$, which means that the braid representation
for this braided surface recovers the full $B_{\ell\nu}(D^2)$. This fact, together
with the existence of Jones unitary representations of the braid groups $B_{\ell\nu}(D^2)$
that span densely the groups $SU(2^N)$ ensures that the braided surface configurations
that arise in geometric models of matter given by Atiyah--LeBrun orbifold edge-cone metrics
suffice to generate a universal quantum computer.  Indeed, it suffices to take the orbifold
edge-cone metric with cone angle $2\pi/3$ and apply the construction  
of \cite{Freedman2} of unitary representations of $B_{3\ell}$ based on the Chern--Simons
TQFT at $5$-th root of unity, which is universal for quantum computation. 

\smallskip

One can then ask, for more general $4$-dimensional 
orbifold geometries $(M,\Sigma)$, where $M$ has a self-dual
Einstein orbifold edge-cone metric near $\Sigma$, whether 
it is always possible to find multisections $S$ 
of the orbifold normal bundle $\cN(\Sigma)$, such that the image of the 
associated braid representation determines a universal quantum computer, 
or whether there are topological and geometric obstructions.  In particular,
one can look for $4$-dimensional geometries related to the models of
matter considered in \cite{AtMan}, \cite{AtMan2}.

\smallskip
\subsection{Additional comments and questions}

The idea of a geometrization of the Skyrmion model originates in the work
\cite{AtMan89}, where Skyrme field configurations in three dimensions 
with a given baryon number $k$ are generated via holonomies 
from $SU(2)$ self-dual Yang--Mills instantons in four dimensions, 
with topological charge $k$. However, the model we discussed in the present
paper is based on a different approach, developed more recently in
\cite{AFS}, \cite{AtMan}, \cite{AMS}, where instead of considering Yang-Mills
instantons in four dimensions, one considers gravitational instantons and
certain more general classes of four-dimensional manifolds (like the algebraic
surfaces considered in \cite{AtMan}) as the
geometric models of Skyrmion-type hadronic physics. There are many 
significant differences between these approaches. For example, the
relation between Skyrme fields and Yang--Mills instantons of \cite{AtMan89}
can be formulated in holographic terms that also provide a mechanism
for the chiral symmetry breaking implemented at domain walls, \cite{Eto}.
In contrast, the geometric models of matter of \cite{AFS}, \cite{AtMan}, 
\cite{AMS} in general do not admit a holographic description: this can
be seen from the fact that these models include cases without boundary
and with positive curvature (like the projective plane), contrary to the 
expected hyperbolicity and codimension one boundary of the holographic
setting. It is still in principle possible that the specific
models we focused on in this paper, which have an embedded surface of
orbifold points, may be suitable for some form of holographic description:
notice however that, unlike the usual holographic setting where the
boundary has real codimension one with respect to the bulk, the setting
we consider would require a more general form of holography
based on a complex codimension one locus, such as a boundary divisor in the
algebro-geometric sense, which is real codimension two in the bulk space.
Investigating a possible approach to holography based on complex codimension
one ``boundary divisors" is
beyond the scope of the present paper, though it represents an interesting
problem in itself. In particular, the anyon states we focus on in this work arise 
from multisections of the orbifold
normal bundle of the embedded surface of orbifold points. We can view the 
associated disk bundle as having locally two complex coordinates, one
on the base Riemann surface and one in the fiber disks in the normal direction.
A hypothetical holographic picture of the type mentioned above may regard 
the role of the complex coordinate in the fiber direction as analogous to a (complexified) 
RG scale, by analogy to the ``holographic renormalization group" interpretation in
AdS/CFT holography, \cite{Fuku}. A more in depth development of this topic would
deserve a separate treatment, but we sketch here briefly an idea of a possible
approach that will be expanded in mode detailed form elsewhere. These
considerations should be regarded as speculative at this stage. The best approach
to a possible boundary/bulk geometry appears to be the one based on tensor networks
(see \cite{Happy} and \cite{EvVid}), which carries with it a natural interpretation of
the additional bulk coordinate as scale parameter with a multiscale entanglement 
renormalization ansatz (MERA), which is in general associated to geometric and
topological properties like triangulations (see for instance \cite{KKR} and \cite{Luo}).
In the setting we have introduced in this paper there is a natural geometric framework
given by the multisections $S$ of the orbifold normal bundle $\cN(\Sigma)$ of the
embedded surface of orbifold points $\Sigma$ inside the $4$-manifold $M$, together
with their branched covering structure $S\to \Sigma$ branched along a finite set of
points $b(S)$. A triangulation or more general decomposition of $\Sigma$ with
vertices at $b(S)$ can be pulled back to consistent triangulations/decompositions
on the multisection $S$, with weights corresponding to the data of the orbifold 
structure. To each such branched cover one would like to associate a tensor
network and a MERA type diagram and RG flow picture that is consistent with
the treatment of anyons in terms of tensor networks as in \cite{Pfe}. We can state
this goal here as an open question and we hope to return to it in future work.

\subsection*{Acknowledgement} The first author received support from the
Clay Mathematical Institute, Trinity College Cambridge, and the University of
Edinburgh. The second author was partially supported by NSF grants 
DMS-1201512, DMS-1707882 and PHY-1205440 and by the Perimeter Institute for Theoretical Physics.  
The second author would also like to thank Andrew Ranicki and
Ida Thompson for their generous hospitality during her visits to the first author in Edinburgh.

\bigskip
\bigskip

\end{document}